\documentclass[a4paper,11pt]{article}
\pdfoutput=1
\usepackage{jcappub}
\usepackage[T1]{fontenc}
\usepackage{amsmath,amssymb,aas_macros}
\usepackage{bm}
\usepackage{multirow}

\title{Small-scale CMB anisotropies induced by the primordial magnetic fields}

\author[a,1]{Teppei Minoda,\note{Corresponding author.}}
\author[a,b]{Kiyotomo Ichiki,}
\author[a]{and Hiroyuki Tashiro}

\affiliation[a]{Department of Physics and Astrophysics, Nagoya University,\\
Chikusa-ku, Nagoya, 464-8602, Japan}

\affiliation[b]{Kobayashi-Maskawa Institute for the Origin of Particles and the Universe,
Nagoya University,\\
Chikusa-ku, Nagoya 464-8602, Japan}

\emailAdd{minoda@nagoya-u.jp}

\abstract{
The primordial magnetic fields~(PMFs) produced in the early universe
are expected to be the origin of the large-scale cosmic magnetic fields.
The PMFs are considered to leave a footprint
on the cosmic microwave background (CMB) anisotropies
due to both the electromagnetic force and gravitational interaction.
In this paper,
we investigate how the PMFs affect the CMB anisotropies
on smaller scales than the mean-free-path of the CMB photons.
We solve the baryon Euler equation with Lorentz force due to the PMFs,
and we show that the vector-type perturbations from the PMFs
induce the CMB anisotropies below the Silk scale as $\ell>3000$.
Based on our calculations,
we put a constraint on the PMFs from the combined CMB temperature anisotropies
obtained by Planck and South Pole Telescope (SPT).
We have found that the highly-resolved temperature anisotropies
of the SPT 2017 bandpowers at $\ell \lesssim 8000$
favor the PMF model with a small scale-dependence.
As a result, the Planck and SPT's joint-analysis puts a constraint
on the PMF strength normalized on the co-moving 1 Mpc scale
as $B_{1\mathrm{Mpc}}<1.5$ nG with Planck and SPT at 95\% C.L.,
while $B_{1\mathrm{Mpc}}<3.2$ nG only with the Planck data at 95\% C.L.
We also discuss the effects on the cosmological parameter estimate
when including the SPT data and CMB anisotropies induced by the PMFs.
}

\begin{document}
\maketitle
\flushbottom

\section{Introduction}
\label{sec:intro}
The origin and evolution of large-scale cosmic magnetic fields
are still open questions in modern cosmology.
Magnetic fields in galaxies are measured to be
$B_\mathrm{gal} \sim 10^{-6}$--$10^{-5}~\mathrm{G}$
through many kinds of observations,
such as the synchrotron radiation, Faraday rotation,
Zeeman effect, and polarization from aligned dust grains
(e.g., a detailed review for magnetic fields in galaxies is Ref.~\cite{2013pss5.book..641B}).
Such $\mu$G magnetic fields in galaxies are considered to be achieved
by the galactic dynamo process related to the galactic rotational motion
\cite{2005PhR...417....1B}.
However, the galactic dynamo process requires a seed field,
and the mechanisms to generate the seed magnetic field remain unknown.
On the other hand,
low-density regions of the Universe, the so-called cosmic voids,
are also magnetized as $B \gtrsim 10^{-18}$--$10^{-15}$ G,
as suggested by observations of $\gamma$-ray and ultra-high-energy cosmic-rays
\cite{2010ApJ...722L..39A, 2012ApJ...747L..14V, 2017PhRvD..96b3010A}.
Such magnetic fields in cosmic voids could be originated
in a cosmological phenomenon in the early universe.

A lot of physical processes in the early universe
predict the generation of magnetic fields, for example,
inflation~\cite{1988PhRvD..37.2743T, 1992ApJ...391L...1R, 1992PhRvD..46.2384T},
phase transition~\cite{1983PhRvL..51.1488H, 1989ApJ...344L..49Q, 1991PhLB..265..258V},
topological defects~\cite{1995PhRvD..51.5946A, 1997MNRAS.287....1S},
and the Harrison mechanism~\cite{1970MNRAS.147..279H}.
Such magnetic fields generated in the early universe
are called the ``Primordial Magnetic Fields'' (PMFs),
and the PMFs can solve both of the above problems;
the origin of the galactic and intergalactic magnetic fields.
Therefore it is important to observationally investigate
the existence and the nature of the PMFs.
Since the PMFs leave different observational signatures,
we can put different constraints on the PMFs
by comparing them with observational data.
We obtain many observational constraints on the PMFs from
the abundance of light elements produced by the Big Bang nucleosynthesis
\cite{1996PhRvD..54.4714C, 2019ApJ...872..172L},
the temperature and polarization anisotropies of the cosmic microwave background (CMB)
\cite{2016A&A...594A..19P, 2017PhRvD..95f3506Z, 2019JCAP...11..028P},
magnetic reheating before the recombination epoch~\cite{2018MNRAS.474L..52S},
the galaxy population~\cite{2016MNRAS.456L..69M},
Lyman alpha data~\cite{2012PhRvD..86d3510S, 2013ApJ...770...47K},
the cluster abundance~\cite{2012MNRAS.424..927T},
and the 21-cm global signal~\cite{2019MNRAS.488.2001M}.

In this paper, we revisit the small-scale CMB anisotropy induced by the PMFs
and give updated constraints on the PMFs from CMB angular power spectra
obtained from the Planck and the latest South Pole Telescope (SPT) data.
Many authors have studied the small-scale CMB anisotropies induced by the PMFs well
\cite{2003MNRAS.344L..31S, 2006PhRvD..73b3002T}.
One of their important features
arises below the photon diffusion scale, the so-called Silk scale.
Although the photon diffusion affects the CMB anisotropies by the PMF,
their power spectrum does not exponentially damp
but decreases in the power-law shape~\cite{1998PhRvL..81.3575S}.
However, in this paper, we will show that
their shape is totally different below the photon mean-free-path scale.
The PMFs can survive even much below the Silk scale
\cite{1998PhRvD..57.3264J, 1998PhRvD..58h3502S}.
Below the mean-free-path scale,
the survived PMFs can accelerate the baryon fluid motions decoupled from the photons
as investigated in Refs.~\cite{1998PhRvL..81.3575S, 2002MNRAS.335L..57S}.
As a result, on such small scales,
although the photon diffusion decreases the CMB anisotropies once,
significant CMB anisotropies can be created again
by the Doppler effect of the baryon fluid velocities driven by the PMFs.
We demonstrate this enhancement on small scales
by solving the baryon Euler equation analytically with simple assumptions.

It is expected that this enhancement provides
the impact on observational constraints on the PMFs.
Nowadays the SPT can reveal the CMB anisotropy angular spectrum
up to $\ell \sim 8000$~\cite{2018ApJ...852...97H} in terms of the multipole~$\ell$,
which corresponds to the scales below the mean-free-path scales.
Therefore we also investigate how much
such small-scale CMB anisotropies affect the constraints on the PMFs
by performing a MCMC analysis.
We provide the first PMF constraint with both temperature and polarization
anisotropies from Planck and SPT including scales below the mean-free-path scale.

This paper is organized as follows:
In section \ref{sec:theory},
we introduce the statistical property of the PMFs
and discuss the CMB angular power spectra sourced by the stress-energy tensor of the PMFs.
Next, we focus on the CMB spectra induced by the PMFs
below the photon mean-free-path scale in section~\ref{sec:calculation}.
Then we describe the method of our MCMC analysis with the Planck and SPT data
that we use to constrain the PMF model parameters in section~\ref{sec:methods}.
We show our new constraint on the PMF strength
and give some uncertainties on the results in section~\ref{sec:results}.
Finally we summarize in section~\ref{sec:conclusion}.

\section{CMB anisotropies created by the PMFs}
\label{sec:theory}

\subsection{Statistical properties of the PMFs}
Since the primordial plasma has high conductivity,
the ideal Magnetohydrodynamics~(MHD) is valid to provide the evolution of the PMFs.
Besides, the back-reaction from the primordial plasma motion on the PMFs
is described as a higher-order effect in the linear cosmological perturbation theory,
and we can neglect them in this paper.
Therefore, the PMF evolution can be described adiabatically.
In these assumptions, following the cosmological expansion,
the PMFs evolve as
$\bm{B}(\bm{x}, t) = \bm{B}_0 (\bm{x}) /a^2(t)$,
where $a(t)$ is the scale factor at time $t$, which is normalized in $a(t_0) =1$
at the present epoch, $t_0$.

Next, we assume that the PMFs are
statistically homogeneous and isotropic Gaussian random fields.
In this case,
the two-point correlation function of the PMFs in Fourier space
can be written with the power spectrum $P_B(k)$ as
\begin{equation}
\langle B_i (\bm{k}) B^*_j (\bm{k}') \rangle
= \delta_\mathrm{D} (\bm{k}-\bm{k}') (\delta_{ij}-\hat{k}_i \hat{k}_j) P_B(k)~.
\label{eq:twopointB}
\end{equation}
Here $\hat{k}_i$ is the $i$-th component of the unit vector of $\bm k$,
and the Fourier component of the PMFs is defined by
$\bm{B}(\bm{k}) = \int d^3 x ~\mathrm{e}^{i\bm{k}\cdot \bm{x}} \bm{B}(\bm{x})$~.
In Eq.~\eqref{eq:twopointB}, we assume no helicity of the PMFs.

We are interested in the magnetic field strength in real space.
Therefore, it is convenient to introduce the
magnetic field strength smoothed on $\lambda =1~\mathrm{Mpc}$ scale, $B_{1\mathrm{Mpc}}$.
This smoothed magnetic field strength is related to
the amplitude of the PMF power spectrum as
\begin{equation}
B_{1\mathrm{Mpc}}^2 = \int \cfrac{d^3 k}{(2\pi)^3}
~\mathrm{e}^{-k^2 \lambda^2} P_B (k)~,
\label{eq:smoothedB}
\end{equation}
where we choose a Gaussian function for the smoothing window function.

For simplicity, we assume that the power spectrum of the PMFs
has a power-law shape as
\begin{equation}
P_B(k) = S_0 (k/k_\mathrm{n})^{n_B}~,
\label{eq:PowerB}
\end{equation}
where $k_\mathrm{n}$ is the wave number for the normalization scale
and we take $k_\mathrm{n} \equiv 2\pi~\mathrm{Mpc}^{-1}$.
By combining equations~\eqref{eq:smoothedB} and~\eqref{eq:PowerB},
we can relate the amplitude of the PMF power spectrum $S_0$
to the smoothed PMF strength $B_\mathrm{1Mpc}$ as
\begin{equation}
B_{1\mathrm{Mpc}}^2 = \cfrac{S_0}{(2\pi)^{n_B+2} (1~\mathrm{Mpc})^3}
~\Gamma\left(\cfrac{n_B+3}{2}\right)~.
\label{eq:relationStoB}
\end{equation}

In this study, we do not focus on any specific PMF generation mechanism
which determines the spectrum of the PMFs.
Instead, we evaluate the CMB anisotropies by taking three free parameters,
$B_\mathrm{1Mpc}, n_B,$ and $\eta_B$ for characterizing our PMF model.
Here, $\eta_B$ is the conformal time at which the PMFs are generated.
Although it does not appears in the power spectrum of the PMFs,
$\eta_B$ is an important parameter in the calculation of the PMFs. 
We will discuss how $\eta_B$ affects the CMB anisotropies later.

The PMFs can create the CMB anisotropies through
generating metric perturbations by their own energy and stress gravitationally
and altering the primordial plasma motion through their electromagnetic interaction.
These effects can be described in the Einstein-Boltzmann equation system
with their energy-momentum tensors.
We can write the energy-momentum tensor of the PMFs in Fourier space,
${T_B}^\mu_{~\nu}$, as
\begin{align}
{T_B}^0_{~0} (\bm{k}, t) &= - \cfrac{1}{8\pi a^4(t)}
\int \cfrac{d^3 k'}{(2\pi)^3} B^a(\bm{k}') B_a(\bm{k}-\bm{k}')
\equiv - \rho_\gamma \Delta_B~,\\
{T_B}^0_{~i} (\bm{k}, t) &= 0~,\\
{T_B}^i_{~j} (\bm{k}, t)
&= \cfrac{1}{4\pi a^4(t)} \int \cfrac{d^3 k'}{(2\pi)^3}
\left[\cfrac{1}{2} B^a(\bm{k}') B_a(\bm{k}-\bm{k}') \delta^i_{~j}
- B^i(\bm{k}') B_j(\bm{k}-\bm{k}')\right]
\nonumber \\
&\equiv p_\gamma (\Delta_B \delta^i_{~j} - {\Pi_B}^i_{~j})~.
\end{align}
Here we have defined the dimensionless parameters $\Delta_B$ and ${\Pi_B}^i_{~j}$
with the energy density $\rho_\gamma$ and pressure $p_\gamma$ of CMB photons.

The scalar-vector-tensor decomposition is powerful to analyze the PMF effect
on the cosmological perturbation quantities including the CMB anisotropies. 
Although $\Delta_B$ is a scalar quantity,
the PMF anisotropic stress tensor ${\Pi_B}^i_{~j}$ can be decomposed
into the scalar $\Pi_B^{(0)}$, vector $\Pi_B^{(\pm 1)}$, and tensor part $\Pi_B^{(\pm 2)}$
(for detailed discussion, see, e.g.,~Refs.
\cite{2002PhRvD..65l3004M, 2010PhRvD..81d3517S}).
Therefore, through the Einstein equation,
$\Delta_B$ and $\Pi_B^{(0)}$ can generate the scalar type of the metric perturbations
and $\Pi_B^{(\pm 1)}$ and  $\Pi_B^{(\pm 2)}$ can create
the vector and tensor types of those, respectively.
The PMFs can source all types of metric perturbations.

Besides the metric perturbation,
the PMFs can alter the motion of charged particles
because of the electromagnetic interaction,
and generate perturbations on the baryon plasma through the Lorentz force.
The Lorentz force term, $L_i$, which appears in the Boltzmann equation of the baryon plasma
(see e.g., Ref.~\cite{2004PhRvD..70d3011L}),
can be related to the magnetic energy density and anisotropic stress.
Thus the Lorentz force term is also decomposed into the scalar-type
$L^{(\mathrm{S})} = 2 \Pi_B^{(0)}/9 - \Delta _B/3$
with the scalar-type values, $\Delta _B$ and $\Pi_B^{(0)}$,
and the vector-type
$L_i^{(\mathrm{V})} = ik(\Pi_B^{(+1)} \hat{k}_{(i} \mathrm{e}_{j)}^+
+ \Pi_B^{(-1)} \hat{k}_{(i} \mathrm{e}_{j)}^-) \hat{k}_j$
with the vector-type value $\Pi_B^{(\pm 1)}$.
Here, $\mathbf{e}^{\pm} \equiv -i/\sqrt{2} (\mathbf{e}^1 \pm i\mathbf{e}^2)$
are the helicity bases with complex orthonormal bases, $\mathbf{e}^1$ and $\mathbf{e}^2$,
which are perpendicular to $\bm{k}$
\cite{2002PhRvD..65l3004M, 2010PhRvD..81d3517S}.

For the complete set of the Einstein-Boltzmann equations
with the PMFs in the cosmological linear perturbation theory,
we refer the readers to Ref.~\cite{2009MNRAS.396..523P}.

\subsection{Impact of the PMFs on the CMB spectra}
In order to study the evolution of the cosmological perturbations,
we solve the Boltzmann and Einstein equations
with some initial conditions including the PMF parameters.
In practice, we set the initial condition on the super-horizon scale
and well after the neutrino decoupling.
Here, due to the impact of neutrino fluid on the metric perturbations,
there are two initial conditions for perturbations created by the PMFs.
After the neutrino decoupling, neutrinos freely stream,
and then they can create non-zero anisotropic stress to compensate that of the PMFs.
On the other hand, before the neutrino decoupling,
neutrinos tightly couple with the photon-baryon fluid.
Therefore, neutrinos have negligible anisotropic stress
and cannot compensate that of PMFs.
This difference between before and after the neutrino decoupling
brings two modes of the perturbations;
passive and compensated modes.

The passive mode arises due to the anisotropic stress of the PMFs
before the neutrino decoupling.
For the scalar and tensor types,
the non-zero anisotropic stress of the PMF
gives rise to the passive mode of the metric perturbations
on both super- and sub-horizon scales soon after the generation of the PMFs.
When neutrinos decouple from the photon-baryon fluid,
the anisotropic stress of the PMFs is compensated
by free-streaming neutrinos, as mentioned before.
The source of the metric perturbations vanishes and the perturbation growth halts after that.
As a result, the induced perturbation can exist
as a constant perturbation on the super-horizon scales,
similarly to an inflationary adiabatic perturbation.
Therefore, the evolution after the horizon entry is the same as 
in case of the inflationary adiabatic perturbation.
The amplitude of these types depends on
$\ln (\eta_\nu/\eta_B)$,
with the conformal time of the neutrino decoupling $\eta_\nu$
and that of the PMF generation
$\eta_B$~\cite{2010PhRvD..81d3517S}.
On the other hand, although the vector-type perturbations
are also generated as the passive mode,
they decay quickly after the neutrino decoupling
as same as the inflationary adiabatic vector-type perturbations do.
Therefore the CMB anisotropies arise
from the scalar and tensor-type perturbation for the passive mode.
The angular power spectrum of the CMB anisotropies due to the passive mode
are proportional to $\langle \Pi_B^{(0)*} \Pi_B^{(0)} \rangle$ for the scalar-type contribution
and $\langle \Pi_B^{(\pm 2)*} \Pi_B^{(\pm 2)} \rangle$ for the tensor-type contribution.

The compensated modes are generated by the PMFs after the neutrino decoupling.
Since the PMF anisotropic stress
is canceled by the neutrino free streaming motion after the neutrino decoupling,
there is no contribution of the PMF anisotropic stress to the metric perturbation.
As a result, the metric perturbations
are not generated at the leading order on super-horizon scales
in the compensated modes.
In this respect, the compensated mode is similar to the isocurvature perturbation.
With no initial perturbation on super-horizon scales,
the PMFs can induce the perturbation sourced by the energy-stress tensors
including Lorentz force acting on the baryon fluid on sub-horizon scales.
Therefore, the compensated mode is important on the small-scale CMB anisotropies.
Although some fraction of the CMB anisotropies due to the compensated modes
is erased by the Silk damping effect,
the CMB anisotropies on smaller scales than the Silk scale
can be created by the compensated mode.
This is because the PMFs can survive and continue to source
the baryon velocity perturbations to create CMB anisotropies
even below the Silk damping scale.
We discuss the behavior of such small-scale CMB anisotropies in section~\ref{sec:calculation}.
The compensated mode includes
the scalar-, vector-, and tensor-type perturbations.
All of them can generate the CMB anisotropies.
In the scalar type,
the amplitude of the CMB angular power spectrum is proportional to
$\langle \Delta_B^{*} \Delta_B \rangle$, $\langle \Pi_B^{(0)*} \Pi_B^{(0)} \rangle$,
and $\langle \Delta_B^{*} \Pi_B^{(0)} \rangle$.
In the vector and tensor types,
$\langle \Pi_B^{(\pm 1)*} \Pi_B^{(\pm 1)} \rangle$
and $\langle \Pi_B^{(\pm 2)*} \Pi_B^{(\pm 2)} \rangle$
appear on the CMB angular power spectrum, respectively.

\begin{figure}
\begin{minipage}{0.5\hsize}
\centering
\includegraphics[width=1.0\linewidth]{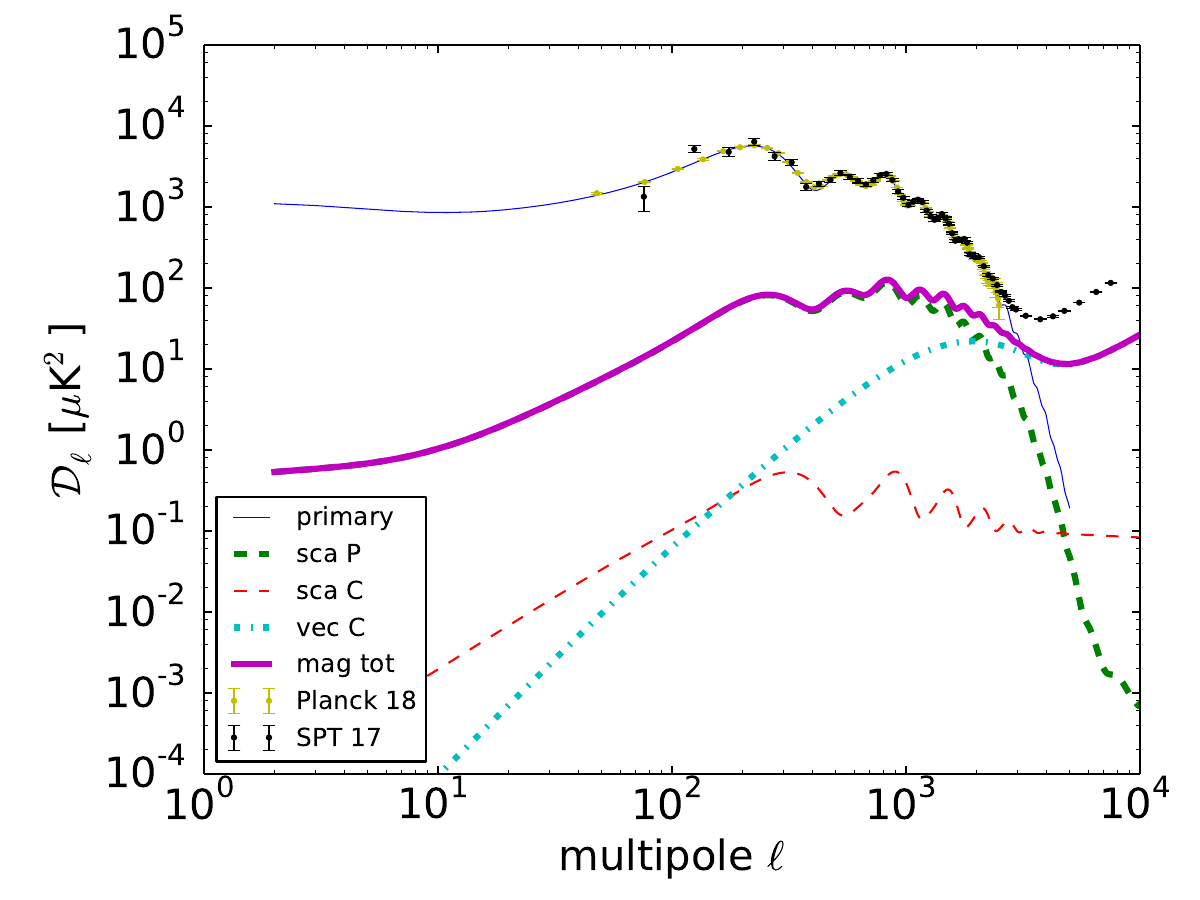}
\end{minipage}
\begin{minipage}{0.5\hsize}
\centering
\includegraphics[width=1.0\linewidth]{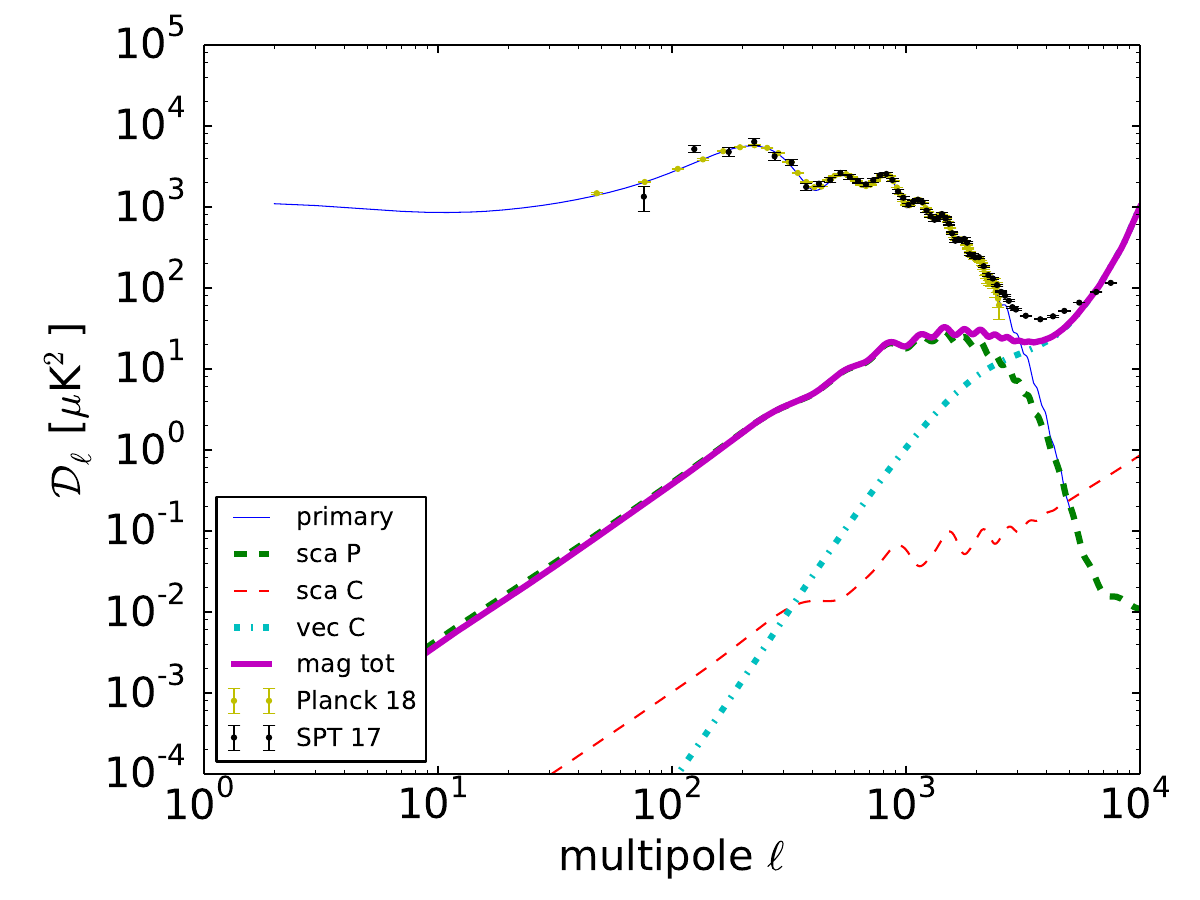}
\end{minipage}
\caption{
The top thin solid line is the primary CMB temperature angular power spectrum,
which is due to the adiabatic perturbation created by the inflation mechanism.
We show the magnetically induced power spectra for three magnetic modes,
the passive scalar mode~(sca P) with the thick dashed line,
the compensated scalar mode~(sca C) with the thin dashed line,
and the compensated vector mode~(vec C) with the dash-dotted line.
In the plots, we take the PMF parameters
as $B_\mathrm{1Mpc}=4.0$ nG and $n_B=-2.5$ for the left panel,
and $B_\mathrm{1Mpc}=3.0$ nG and $n_B=-1.0$ for the right panel.
The compensated scalar and tensor modes are not shown
because they have relatively smaller amplitudes on large multipoles.
We also show the total magnetic contribution as the thick solid line.
}
\label{fig:CMB}
\end{figure}

In Fig.~\ref{fig:CMB}, we summarize
the CMB temperature-temperature (TT) auto-power spectra
induced by the PMFs for three dominant contributions, namely,
the passive and compensated scalar modes, and the compensated vector mode.
We neglect the tensor-type contribution
because the created CMB anisotropies are much smaller
than the other types contributions on small scales.
(e.g., see Fig.~2 in Ref.~\cite{2010PhRvD..81d3517S}).
It is clearly seen that the compensated vector mode is dominant for $\ell \gtrsim 4000$.
When plotting Fig.~\ref{fig:CMB}, we fix the PMF generation epoch
at the grand unified theory (GUT) phase transition, as $\eta_\nu/\eta_B = 10^{17}$.
We also plot the measurements of the CMB temperature anisotropies
obtained by Planck and SPT in Fig.~\ref{fig:CMB}.
The comparison with the current observation status tells us that
small-scale CMB anisotropies due to the compensated vector perturbations are crucial
to obtain the constraint on the PMFs
because the compensated vector perturbations have
a significant blue-tilted angular power spectrum on small scales.
In the next section,
we represent how such a blue-tilted spectrum is created by
the compensated vector perturbations.

\section{The compensated vector mode on small scales}
\label{sec:calculation}
\begin{figure}
\centering
\includegraphics[width=0.8\hsize]{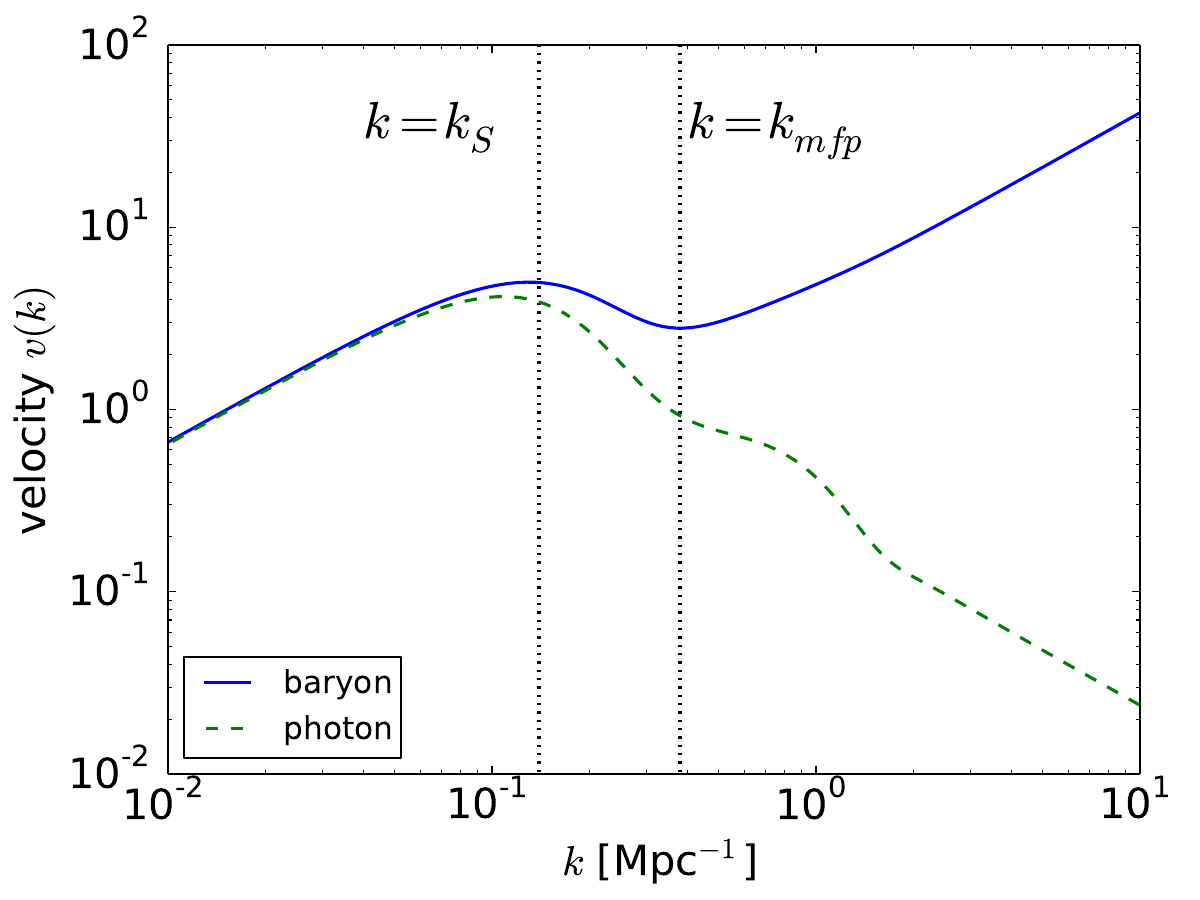}
\caption{
The scale-dependent baryon velocity perturbation and
the photon dipole moment at the recombination are shown
by the solid and dashed lines, respectively.
The model parameters of the PMFs are fixed with
$B_\mathrm{1Mpc}=3.0$ nG and $n_B=-1.0$.
}
\label{fig:vector}
\end{figure}

In the compensated vector mode,
the PMF can create the perturbations of the divergence-free baryon velocity and metric.
Therefore, the CMB temperature anisotropies are created from
the Doppler and integrated Sachs-Wolfe (ISW) effect
with these perturbations~\cite{1994FCPh...15..209D}.
The observable temperature anisotropies can be written as
\begin{align}
\Theta (\eta_0,k, \hat{\bm{n}})
&=\int^{\eta_0}_{0} \mathrm{e}^{-\tau(\eta_0,\eta)}
[{\tau}'(\eta) \bm{v}_\mathrm{b}(\eta, k) - {\bm{V}}'(\eta, k)]
\cdot \hat{\bm{n}} ~d\eta \nonumber \\
&=\int^{\eta_0}_{0}~g(\eta_0,\eta)
\bm{v}_\mathrm{b}(\eta, k) \cdot \hat{\bm{n}}~d\eta
+\int^{\eta_0}_{0} \mathrm{e}^{-\tau(\eta_0,\eta)}
{\bm{V}}'(\eta, k) \cdot \hat{\bm{n}}~ d\eta~,
\label{eq:theta_vector}
\end{align}
where
$\hat{\bm{n}}$ is a line-of-sight unit vector,
and $\eta_0$ denotes the present time.
The dotted quantities represent derivatives with respect to the conformal time $\eta$.
We take the photon opacity as $\tau' \equiv an_e \sigma_\mathrm{T}$,
the optical depth as
$\tau(\eta_1, \eta_2) = \int^{\eta_1}_{\eta_2} \tau' d\eta$,
and the visibility function as
$g(\eta_0,\eta) \equiv {\tau}'(\eta) \mathrm{e}^{-\tau(\eta_0,\eta)}$.
Additionally, $\bm{v}_\mathrm{b}(\eta)$ is divergence-free baryon velocity perturbation,
and $\bm{V}(\eta)$ is the gauge-invariant metric perturbation
in vector mode as defined in Ref.~\cite{1980PhRvD..22.1882B}.
\footnote{By using the integration by parts and the spherical harmonics expansion,
Eq.~\eqref{eq:theta_vector} gives Eq.~(1) in Ref.~\cite{2002MNRAS.335L..57S}.}
In order to calculate the CMB anisotropies from the compensated vector mode,
we need to solve
the Euler equation of baryons for $\bm{v}_\mathrm{b}(\eta)$
and Einstein equation for $\bm{V}(\eta)$
\footnote{In general, besides them, we also need to take into account 
the Boltzmann equations of photons, dark matter, and neutrinos.
However, the impact of these components is negligible in our interested case.
Therefore we do not consider their evolutions.}.
However, we do not discuss the evolution of the vector potential $\bm{V}(\eta)$
because its contribution to the CMB anisotropies
via the ISW effect is negligible on small scales.
In other words, the enhancement of the CMB anisotropies on small scales
comes from the scale-dependence of the baryon velocity at the recombination,
$\bm{v}_\mathrm{b}(\eta_\mathrm{rec}, k)$.
In order to confirm this statement,
we plot the fully numerical solution for $\bm{v}_\mathrm{b}(\eta_\mathrm{rec}, k)$
calculated by the Boltzmann code~\cite{2017PhRvD..95f3506Z} in Fig.~\ref{fig:vector}.
This shows that the scale dependence of the baryon velocity
can be divided into three different scales:
the large scale $k \lesssim k_S \approx 0.14$ Mpc$^{-1}$,
the intermediate scale
$k_S \lesssim k \lesssim k_\mathrm{mfp} \approx 0.38~$Mpc$^{-1}$,
and the small scale $k \gtrsim k_\mathrm{mfp}$.
Here $k_S$ and $k_\mathrm{mfp}$ denote
the Silk damping scale and the mean-free-path scale of the CMB
at the recombination epoch, respectively.

For large and intermediate scales as $k \lesssim k_\mathrm{mfp}$,
the evolution of the vector perturbation has been
analytically studied with the tight-coupling approximation,
$\bm{v}_\mathrm{b} \simeq \bm{v}_\gamma$,
in Refs.~\cite{1997PhRvD..56..596H, 2002PhRvD..65l3004M}.
According to their results,
for the larger scales than the Silk scale
(or on the earlier phase of the evolution),
although the Lorentz force acts on the baryon fluid and increases the baryon velocity,
the photon velocity~(i.e., dipole moment, $\bm{v}_\gamma$)
catches up baryons quickly because of the tight coupling due to the Compton scattering.
As a result, the baryon velocity and the photon dipole moment
are growing with wavenumber (or time) together
for $k \lesssim k_S \approx 0.1$ Mpc$^{-1}$.
However, for the intermediate scales,
the effect of non-zero mean-free-path for photons is not negligible.
The non-zero mean-free-path loses the tight coupling and 
affects the perturbation evolution as the photon viscosity 
in the photon-baryon plasma.
As a result, 
the baryon and photon velocity perturbations, 
$\bm{v}_\mathrm{b}$
and $\bm{v}_\gamma$,
starts to decay.

As shown in Fig.~\ref{fig:vector},
there exists a weak coupling between baryons and photons below the Silk scale
and they continue to decrease for a while.
However, when the photon mean-free-path scale is larger than the perturbation scale,
the baryon-photon coupling no longer holds.
After the decoupling, the baryon velocity starts to grow due to the PMFs again.
Now let us obtain the analytical expression of the baryon velocity evolution on such small scales.
Below the Silk scale, the photon dipole moment becomes small more quickly than the baryon velocity.
Neglecting the photon dipole moment,
we can write the Euler equation of the baryon fluid in the presence of the PMFs in
\begin{equation}
\left(\frac{\partial}{\partial t} + H + \cfrac{\tau'}{R} \right)
\bm{v}_\mathrm{b}
= -\cfrac{\rho_\gamma \bm{L}^{(V)}}{2 a \rho_\mathrm{b}}~,
\label{eq:EoMforBaryon}
\end{equation}
in the non-relativistic limit
\cite{1998PhRvD..57.3264J, 2004PhRvD..70l3003B}.
Here $\rho_\mathrm{b}$ and $\rho_\gamma$
are the energy densities of baryons and photons respectively,
and $R \equiv 3\rho_\mathrm{b}/4\rho_\gamma$
is the total energy and pressure ratio of baryons and photons.
The left-hand side of equation~\eqref{eq:EoMforBaryon}
includes the Hubble expansion, Compton scattering,
and the right-hand side represents the Lorentz force due to the PMFs.
Before the recombination epoch, we can neglect the Hubble expansion term, in comparison with the Compston scattering term.
Solving equation~\eqref{eq:EoMforBaryon},
we obtain
\begin{align}
\bm{v}_\mathrm{b} (k, a) =
& \bm{v}_\mathrm{b,i} \exp\left\{\cfrac{\tau'}{2aHR}
\left[1-y_\mathrm{i}^2(a) \right] \right\} \nonumber \\
&-\cfrac{\rho_\gamma \bm{L}^{(V)}}{4aH\rho_\mathrm{b}}
\left[\operatorname{Ei} \left(-\left.\cfrac{\tau'}{2aHR}\right|_\mathrm{i}\right)
-\operatorname{Ei} \left(-\cfrac{\tau'}{2aHR}\right)\right]
\exp \left(\cfrac{\tau'}{2aHR}\right), \nonumber \\
&\hspace{86mm} (a \lesssim a_\mathrm{rec}),
\label{eq:SmallScaleSolution-before}
\end{align}
where $\operatorname{Ei}(x)\equiv \int^x_{-\infty} e^t/t~dt$ is the exponential integral,
$y_\mathrm{i} (a)\equiv a/a_\mathrm{i}$ and $y_\mathrm{rec} (a)\equiv a/a_\mathrm{rec}$ are
the scale factors normalized at the initial time and at the recombination time,
and the values with the subscript ``i'' and ``rec'' represent
the ones at the initial time and at the recombination time, respectively.

\begin{figure}
\centering
\includegraphics[width=0.8\hsize]{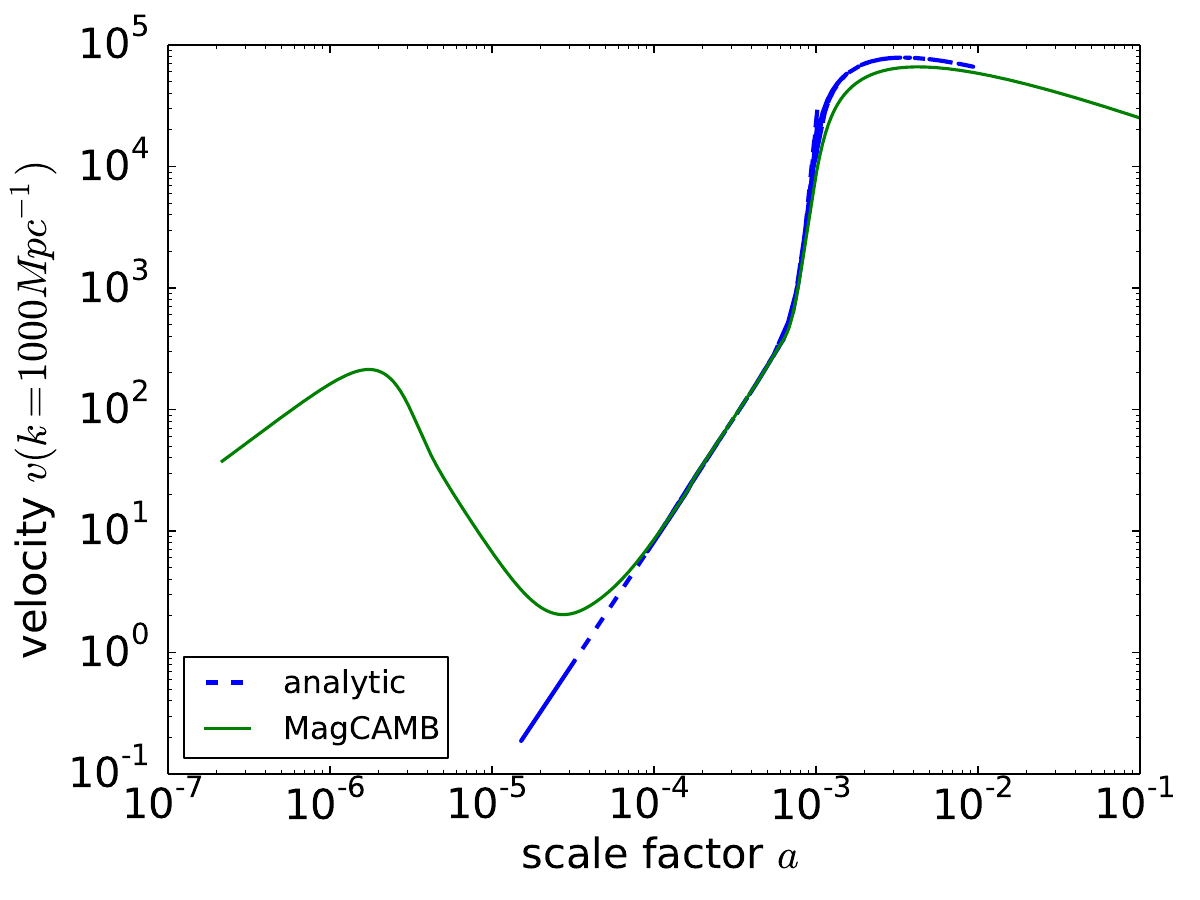}
\caption{
Numerical solution for the baryon velocity and
the analytical solution that we have obtained.
After the baryon velocity damps due to the photon viscous effect,
our analytical solution is in good agreement with the numerical solution
for $10^{-5} \lesssim a \lesssim 10^{-3}$.
Here we fix $k=1000$ Mpc$^{-1}$ and
the PMF parameters are the same as in Fig.~\ref{fig:vector}.
}
\label{fig:baryon_velocity}
\end{figure}

Around the recombination epoch, the Hubble expansion term starts to dominate the Compton scattering term in equation~\eqref{eq:EoMforBaryon}.
Therefore, neglecting the Compton scattering term,
we get
\begin{align}
\bm{v}_\mathrm{b} (k, a) =
& \bm{v}_\mathrm{b,rec}~y_\mathrm{rec}^{-1} (a)
+ \left(\left.\cfrac{\rho_\gamma \bm{L}^{(V)}}{H\rho_\mathrm{b}}\right|_\mathrm{rec}
- \cfrac{\rho_\gamma \bm{L}^{(V)}}{H\rho_\mathrm{b}} \right) a^{-1},
\hspace{13mm} (a \gtrsim a_\mathrm{rec})~.
\label{eq:SmallScaleSolution-after}
\end{align}
In equations~\eqref{eq:SmallScaleSolution-before} and
\eqref{eq:SmallScaleSolution-after},
the first terms on the right hand side
represents the decaying terms
due to the Compoton scattering and
the Hubble expansion, respectively. 
On the other hand,
the second terms in both equations are the inhomogeneous solution
which comes from the PMF source term,
i.e., from the right hand side in equation~\eqref{eq:EoMforBaryon}.
In Fig.~\ref{fig:baryon_velocity},
we plot the evolutions of the baryon velocity fluctuation from 
equations~\eqref{eq:SmallScaleSolution-before}
and~\eqref{eq:SmallScaleSolution-after}
in the dashed line
and one from the full numerical calculation in the solid line.
Here we set the initial time at $a_\mathrm{i}=10^{-5}$.
The figure tells us that
our solutions agree with the full numerical calculation
in particular before the recombination epoch.

CMB temperature anisotropies on small scales are 
created by the baryon velocity perturbation
due to the Doppler effect in the recombination epoch as discussed earlier.
Through this section, we have obtained the time evolution
of $\bm{v}_\mathrm{b}$ on smaller scales than the photon mean-free-path scale,
and we have shown that the second term in equation~\eqref{eq:SmallScaleSolution-before}
is dominant around the recombination epoch.
As a result, the $k$-dependence of
$|\bm{v}_\mathrm{b}|_\mathrm{rec}^2$ is given by
\begin{align}
|{\bm v}_b |^2_{\rm rec} \propto |\bm{L}^{(V) }|^2 \propto k^2 \Pi^2(k)~.
\end{align}
The enhanced baryon velocity on small scales
induces the high-$\ell$ CMB temperature anisotropies
as shown in Fig.~\ref{fig:CMB}.

\section{Data Analysis with Planck and SPT}
\label{sec:methods}

The aim of this paper is to obtain the constraint on the PMFs from the new small-scale CMB anisotropies obtained by SPT.
We use the Markov Chain Monte Carlo (MCMC) method
in order to provide the constraint on the PMF model parameters, $(B_{\rm 1Mpc}, n_B, \tau_B)$.
For the MCMC analysis, we use publicly available numerical codes,
\texttt{MagCAMB} and \texttt{MagCosmoMC},
developed by Zucca et al.~\cite{2017PhRvD..95f3506Z}.
\texttt{MagCAMB} is the code to solve the
linearized Boltzmann and Einstein equations with the PMFs.
Based on \texttt{CAMB}~\cite{Lewis:1999bs}
and its modified version~\cite{2010PhRvD..81d3517S},
\texttt{MagCAMB} can calculate the CMB angular power spectra
for all contributions sourced by the PMFs mentioned in the previous section.
\texttt{MagCosmoMC} is developed,
based on \texttt{CosmoMC}~\cite{2002PhRvD..66j3511L}.
It enables us to explore
cosmological parameters and the PMF model parameters
with the foreground and data calibration parameters.

We investigate
how the small-scale measurement of the CMB anisotropy by SPT
improves the constraint on the PMF parameters,
compared to the previous constraints by Planck data
\cite{2016A&A...594A..19P, 2017PhRvD..95f3506Z}.
However, it is difficult to constraint the standard cosmological parameters because they are well constrained by the CMB data on larger scales than the SPT scales.
Therefore, we use the combined data
for Planck 2015 and SPTpol 2015/2017.
Planck 2015 data includes
the low-$\ell$ ($2 \le \ell \le 29$) TT, EE, TE, and BB power spectra data,
and high-$\ell$ ($30 \le \ell \le 2508$) TT, TE, and EE power spectra
\cite{2016A&A...594A..11P}.
SPTpol 2015 data includes the BB power spectra
in $300 \le \ell \le 2300$ for three spectral combinations,
95 GHz $\times$ 95 GHz, 95 GHz $\times$ 150 GHz,
and 150 GHz $\times$ 150 GHz~\cite{2015ApJ...807..151K},
and SPTpol 2017 data contains
the TT, TE, and EE power spectra in $50 \le \ell \le 8000$
for the frequency band, 150 GHz $\times$ 150 GHz~\cite{2018ApJ...852...97H}.
We note that Zucca et al.~\cite{2017PhRvD..95f3506Z} has already studied
the constraint on the PMFs with the Planck and SPTpol 2015 BB bandpowers.
In this work, we update the constraint on the PMFs
by adding small-scale TT, TE, and EE data from SPTpol 2017.

One problem is that the SPT 2017 bandpowers are partially
overlapped with the Planck power spectra for $50 \le \ell \le 2508$,
which corresponds to acoustic oscillation scales.
We used both Planck and SPT for these multipole scales,
and this possibly leads to erroneously tighter constraints on the PMF parameters.
However, this acoustic oscillation region is mainly determined
by the standard cosmological parameters,
and we can expect that the constraint on the PMF parameters
is not affected much by using the duplicated CMB data on these scales.

We show the CMB temperature anisotropies
measured by Planck and SPT in Fig.~\ref{fig:CMB}.
The PMF contributions to the CMB angular power spectra
with $B_\mathrm{1Mpc} = 4.0~\text{nG}, n_B = -2.5$ are plotted in the left panel,
and those with $B_\mathrm{1Mpc} = 3.0~\text{nG}, n_B = -1.5$
are shown in the right panel.
Both of these parameter combinations are allowed by the previous constraint
from Planck collaboration~\cite{2016A&A...594A..19P}
with a 95\% confidence level (C.L.).
However, we expect that the high-$\ell$ CMB spectrum measured by SPT
can constrain large $n_B$ as shown in Fig.~\ref{fig:CMB}.
Clearly, the CMB spectra observed by SPT
have tiny errors on small-scales as $\ell \gtrsim 2000$,
and the effect on such small-scale CMB anisotropy
is dominated by the compensated vector mode,
which is relevant for the parameter estimation using the SPT data.
Finally, the blue-tilted PMF spectra
(this corresponds to a large spectral index $n_B$)
that enhance the small-scale CMB anisotropies
are tightly constrained by the high-$\ell$ CMB measurement of SPT.

\section{Results and Discussion}
\label{sec:results}
\begin{figure}
\centering
\includegraphics[width=0.5\hsize]{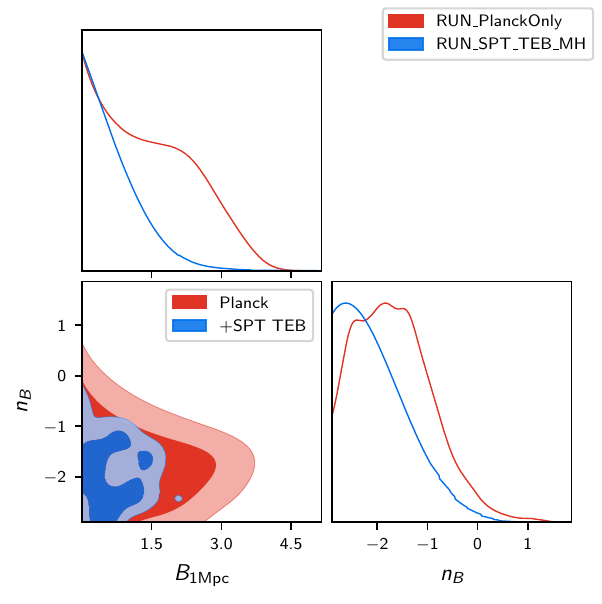}
\caption{
The constraint on the magnetic field strength $B_{1 \mathrm{Mpc}}$
and the spectral index of the PMFs $n_B$.
The thick and thin color region stands for 68\% and 95\% confidence level (C.L.),
respectively.
}
\label{fig:mag2d}
\end{figure}

First, we show the constraint on the PMF parameters, ($B_{1 \mathrm{Mpc}}$, $n_B$),
derived from {\ttfamily MagCosmoMC}
with the Planck and SPT data in Fig.~\ref{fig:mag2d}.
For comparison, we plot the constraint with only the Planck data,
which is the same analysis as done in the previous works
\cite{2016A&A...594A..19P, 2017PhRvD..95f3506Z}.
We have found the normalized PMF strength and the PMF spectral index
can be constrained more tightly with the SPT data.
For the constraint with Planck and SPT,
the MCMC analysis consumes computing time
dozens of times longer than for the Planck only.
This is because the calculation
includes the CMB anisotropies up to $\ell \sim 8000$ and,
on small scales, there arises a strong degeneracy
between the PMF and foreground parameters.
In Fig.~\ref{fig:mag2d}, our analysis with the Planck and SPT
does not reach the convergence of the MCMC analysis.
As one of the most popular methods to evaluate the convergence,
the MCMC analyses usually impose
the so-called Gelman-Rubin diagnostic, $R-1<0.01$,
with $R$ being the square root of the ratio of
the marginalized variance of all chains and that for each chain.
Our analysis with the Planck and SPT in Fig.~\ref{fig:mag2d} finds $R>10$.
Therefore the 2D color contour in Fig.~\ref{fig:mag2d} is not a smooth shape.
However, the marginalized one-dimensional constraints on $B_{1 \mathrm{Mpc}}$ and $n_B$
seem to have smooth probability distribution functions in Fig.~\ref{fig:mag2d}.
Besides, we found that, after reaching the current level of the constraint
in Fig.~\ref{fig:mag2d},
the upper limit on the PMF parameters does not change
for a long time during the calculation.
Therefore we conclude that
the constraint on the PMF parameters in Fig.~\ref{fig:mag2d}
is not far from the result that would be obtained from the converged MCMC analysis.

The upper limits for $B_\mathrm{1Mpc}$ and the best-fitted values for $\tau_B$
from Planck and SPT data are presented in Table~\ref{tab:best-fit}.
The PMF amplitude smoothed on 1 Mpc is constrained as
$B_\mathrm{1Mpc}<1.52$ nG for a 95 \% confidence level (C.L.)
when we include the SPT bandpowers for MCMC analysis,
while the Planck 2015 data only put the upper limit
as $B_\mathrm{1Mpc}<3.18$ nG.
On the other hand, the change on $\eta_{\nu}/\eta_B$ mainly
affects the amplitudes of the passive mode perturbation,
and consequently, it influences the large-scale CMB anisotropies.
Therefore we have an almost unchanged constraint on the $\eta_{\nu}/\eta_B$
when adding the small-scale SPT data for the parameter estimation.
In Table~\ref{tab:best-fit},
although we put the best-fitted value for $\eta_{\nu}/\eta_B$,
we do not show the error-bars for the Planck and SPT case.
This is because we have used duplicated information
of CMB angular power spectra for $50 \le \ell \le 2500$,
and may we underestimate the error, as explained in section \ref{sec:methods}.

\begin{table}
\centering
\caption{
The best-fitted values for the epoch of the PMF generation, $\eta_B$,
and the upper limits for $B_\mathrm{1Mpc}$ with 95\% C.L.
only from the Planck data, and those from the Planck and SPT data are shown.
}
\label{tab:best-fit}
\vspace{5mm}
\begingroup
\renewcommand{\arraystretch}{1.2}
\begin{tabular}{|l||c|c|}
\hline
\multirow{2}{*}{Parameters}
& \multicolumn{2}{c|}{PMF} \\
\cline{2-3}
& Planck & Planck + SPT
\\ \hline
$B_\mathrm{1Mpc}$ & $<$ 3.182 & $<$ 1.515 \\ \hline
$\log \left(\eta_\nu/\eta_B\right)$
& $9.92^{+2.09}_{-5.92}$ & 9.41 \\ \hline
\end{tabular}
\endgroup
\end{table}

\begin{figure}
\centering
\includegraphics[width=\textwidth]{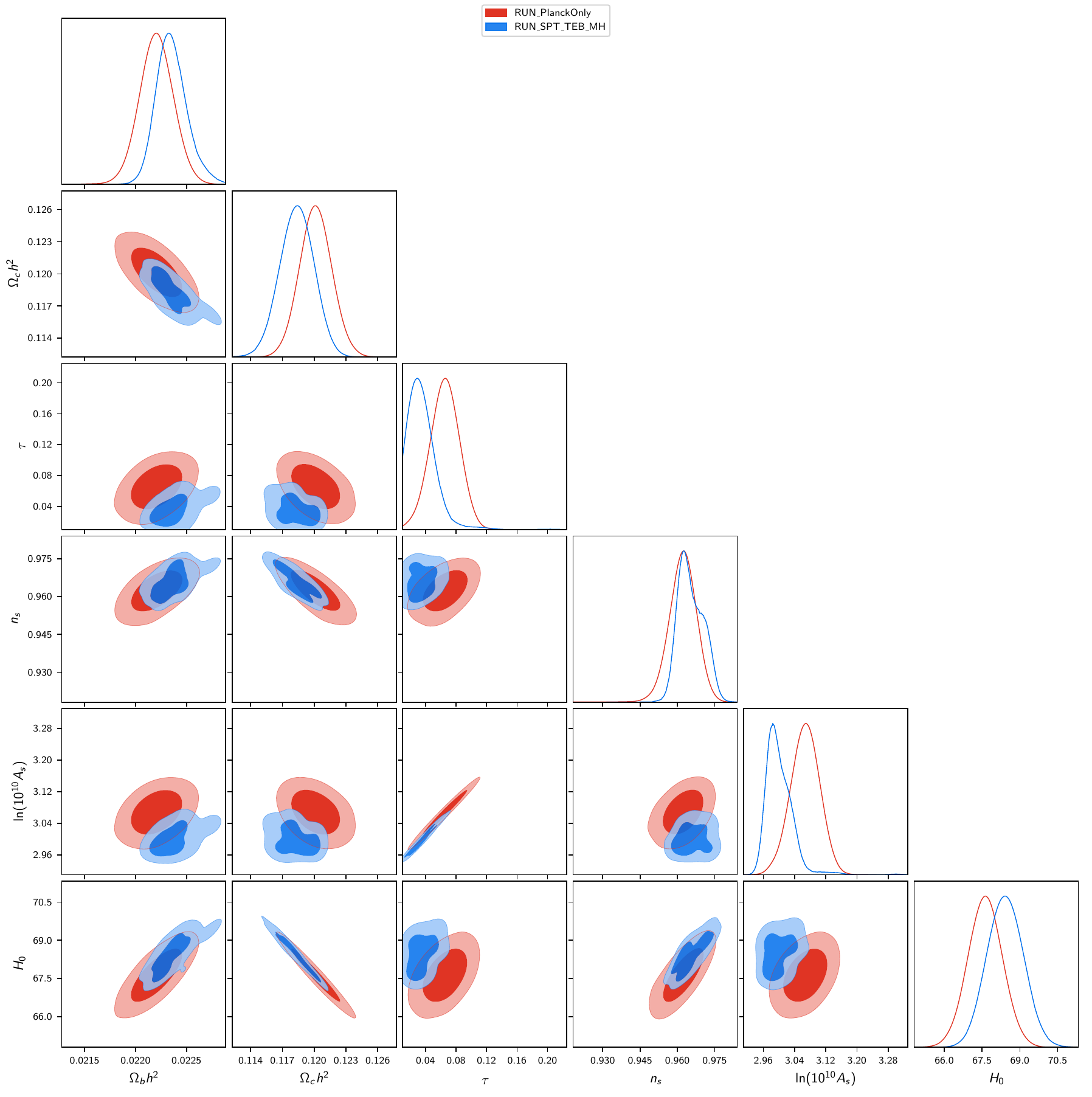}
\caption{
The constraint on the cosmological parameters obtained by MCMC analysis with the PMFs.
The thick and thin color region stands for 68\% and 95\% confidence level, respectively.
The red (RUN-PlanckOnly) and blue (RUN-SPT-TEB-MH) contours indicate
the constraints from Planck 2015 data and those from both of Planck and SPT data, respectively.
}
\label{fig:params_tri}
\end{figure}

Next, we discuss the impact of the PMF parameter on
the determination of the $\Lambda$CDM cosmological parameter.
In Fig.~\ref{fig:params_tri},
we compare the $\Lambda$CDM cosmological parameter constraint
with only Planck data and
that with Planck and SPT TT/TE/EE, and BB data,
including the PMF parameters.
Our results indicate that
the cosmological parameter estimation can be biased
when considering the PMF effects and the SPT high-$\ell$ data.
In particular, Thomson optical depth $\tau$
and the amplitude of the primordial power spectrum for the scalar sector $A_s$
decrease by 1-2 $\sigma$ compared with the Planck only analysis.
The SPT TT data favor a nearly scale-invariant spectrum of the PMFs
as shown in Fig.~\ref{fig:mag2d}.
In this case, the passive tensor mode significantly
enhances the low-$\ell$ EE power spectrum.
To compensate for the enhancement on large scales by the PMFs,
the MCMC chooses the small optical depth $\tau$,
which also contributes to the signals on large scales of the CMB EE power spectrum.

As dedicated in section~\ref{sec:methods},
the measured ranges of the TT/TE/EE spectra which we have used
are overlapped for Planck and SPT for $50 \lesssim \ell \lesssim 2500$.
However, this is not harmful to our results
because the CMB spectra for these angular scales
are almost determined only with the cosmological parameters,
and are not affected by the PMF contribution very much.
Actually, we have confirmed that
the best-fitted values of the cosmological parameters
are not changed for Planck and Planck+SPT,
when excluding the PMFs for our analyses ($\Lambda$CDM case).
For similar reasons, we do not perform BICEP2/Keck-Planck joint analysis
\cite{2016A&A...594A..19P, 2017PhRvD..95f3506Z}.

Before closing this section, we make some comments
on other PMF effects on the CMB anisotropies,
which are not included in our analysis.
Previous studies predict the generation of the helical PMFs during inflation
\cite{2017JCAP...12..002K, 2018PhRvD..97h3503S,
2019JCAP...09..008F, 2020JCAP...01..043P}.
While such helical PMFs can induce the parity-odd spectrum, TB and EB,
the contributions of the helical part of the PMFs
to the parity-even TT, TE, EE, and BB power spectra are always subdominant
in comparison with the contribution
of the non-helical PMFs~\cite{2016A&A...594A..19P}.
Therefore,
adding the helical component of the PMFs to parameter estimation
would not significantly change our constraint.
The dissipation of magnetic fields before the recombination epoch
affects the thermal history and recombination history of the baryon gas
\cite{1998PhRvD..57.3264J, 1998PhRvD..58h3502S,
2005MNRAS.356..778S, 2015MNRAS.451.2244C}.
Including this effect can improve the constraint on the PMFs
via increasing the energy density of CMB photons~\cite{2018MNRAS.474L..52S},
creating $y$-type distortion~\cite{2000PhRvL..85..700J, 2014JCAP...01..009K},
and changing the evolution of Thomson optical depth~\cite{2015JCAP...06..027K}.
Including these effects can improve our constraint on the PMFs,
in particular, with a large spectral index.
However, the dissipation of the PMFs is a strong nonlinear process,
and there is still a large uncertainty in the calculation of these effects.
Therefore we do not include the dissipation effect in our analysis.
The PMFs also can induce the non-Gaussian CMB anisotropies
\cite{2010PhRvD..82l1302S, 2011PhRvD..83l3003S}.
The current constraint on the non-Gaussianity
in the CMB anisotropies by Planck observation
provides the same order of our PMF constraint~\cite{2016A&A...594A..19P}.
Moreover, the CMB polarization map should be altered
by Faraday rotation, if the PMFs exist.
The upper limit on the PMF amplitude via Faraday rotation
from the Planck observation is much weaker than
via the other effects described here.
The PMFs can also create a secondary contribution
to the CMB temperature anisotropy,
which is the so-called thermal Sunyaev-Zel'dovich (tSZ) effect
\cite{2017PhRvD..96l3525M}.
This effect enhances the small-scale CMB anisotropy
of $\ell^2 C_\ell \sim 10-100$ $\mu$K$^2$
at $\ell \sim 10^6-10^7$ for sub-nano Gauss PMFs.
Of course, the CMB anisotropy on such scales cannot be measured
even with the present and on-going radio telescope
including CMB-S4~\cite{2016arXiv161002743A, 2019arXiv190704473A}.
Therefore, we have not included
the tSZ power spectrum from the PMFs into our analysis.

\section{Summary}
\label{sec:conclusion}
In this paper, we have investigated
the impact of the PMFs on small-scale CMB anisotropies.
Then we have provided the constraint on the PMFs
from the CMB temperature and polarization anisotropies
observed by the latest Planck and SPT data.
If the PMFs exist, the stress-energy tensor of the PMFs
induces the additional metric perturbation besides the primordial curvature perturbation,
and the Lorentz force acts on the motion of primordial plasma.
In the analysis,
we have assumed that the PMFs have a simple power-law spectrum,
as $P_B(k) \propto B_{1 \mathrm{Mpc}}^2 k^{n_B}$.
We have also taken the assumption of the ideal MHD approximation for the PMF evolution
and neglected the helicity of the PMFs.
Under these assumptions,
we have performed a numerical calculation
to solve the Einstein-Boltzmann equation system.
To obtain the CMB anisotropy angular power spectra with the PMFs,
we consider
the time evolution of the scalar and tensor perturbations for the passive mode,
and the scalar, vector, and tensor perturbations for the compensated mode.
While the primary CMB anisotropies are rapidly suppressed
for $\ell \gtrsim 1000$ due to the Silk damping effect,
the compensated vector perturbation from a blue-tilted PMFs
significantly contributes to the small-scale CMB anisotropies
as we have shown in Fig.~\ref{fig:CMB}.
In this paper, we have given an analytical solution for
the Euler equation in the small-scale limit for the first time.
We also have shown that the derived baryon velocity perturbation
induces the dominant CMB anisotropies on $\ell \gtrsim 4000$,
depending on the spectral indices of the PMFs.
Therefore, we have found that the nearly scale-invariant PMFs are favored
by the high-resolution measurement of the CMB anisotropies with SPT.

As a result, we have obtained the constraint on the PMFs
as $B_\mathrm{1Mpc}<1.52$ nG for 95 \% C.L.
with the Planck 2015 and SPT 2015/2017 bandpowers.
Our analysis improves the PMF constraint compared with the Planck 2015 constraint.
We have found that the cosmological parameter estimation is biased
when including the PMF parameters and SPT data.
The PMF induced perturbation with the small spectral index $n_B$
enhances the low-$\ell$ polarization anisotropies,
and this results in the smaller Thomson optical depth $\tau$
and the primordial scalar power spectrum amplitude $A_s$.

Finally, we mention the rigidity and future outlook about our constraint.
First, throughout this paper,
we assume that the PMFs are frozen in, and they evolve adiabatically.
However, small scale velocity perturbations are reduced by the Compton scattering
as discussed in section~\ref{sec:calculation}.
Therefore the PMFs also might be suppressed to some degree
by the back-reaction from such radiative diffusion of baryon perturbations.
For instance, a full MHD simulation probably give a more accurate speculation
into the cosmological perturbations and CMB anisotropies
including such non-linear effects.
Next, in addition to the baryon velocity perturbation as shown in this paper,
small-scale PMFs can fluctuate the baryon density fields.
These inhomogeneities can affect the recombination history,
and lead tighter constraints from the CMB anisotropies
\cite{2013JCAP...10..050J, 2019PhRvL.123b1301J}.
Moreover, on small scales,
the degeneracy between the PMF and foreground parameters arises
and brings an impact on the constraint on the PMFs.
Subtracting the foreground is important not only in the CMB constraint on the PMFs
but also in understanding galaxy cluster physics
through the SZ effect~\cite{1970Ap&SS...7....3S},
the CMB gravitational lensing~\cite{1987A&A...184....1B},
and the reionization process with the CMB Doppler effect
in the patchy reionization~\cite{1986ApJ...306L..51O}.
In principle, multi-frequency observation
might help to remove the CMB foreground and improve our constraint.
However, this is beyond the scope of this paper,
and we put these issues on the future works.

\acknowledgments
This work is based on observations obtained with Planck (http://www.esa.int/Planck),
an ESA science mission with instruments
and contributions directly funded by ESA Member States, NASA, and Canada.
Numerical computations were in part carried out on Cray XC50
at Center for Computational Astrophysics, National Astronomical Observatory of Japan.
This work was supported by JSPS (Japan Society for the Promotion of Science)
KAKENHI Grant Number 19J13898 (TM), 18K03616 (KI), and 17H01110 (HT).

\appendix

\bibliographystyle {JHEP}
\bibliography{paper}

\end{document}